**Evidence-based gene models for structural and functional annotations of the oil palm genome**

Chan Kuang Lim[1,2#], Tatiana V. Tatarinova[3#], Rozana Rosli[1,4], Nadzirah Amiruddin[1], Norazah Azizi[1], Mohd Amin Ab Halim[1], Nik Shazana Nik Mohd Sanusi[1], Jayanthi Nagappan[1], Petr Ponomarenko[3], Martin Triska[3,4], Victor Solovyev[5], Mohd Firdaus-Raih[2], Ravigadevi Sambanthamurthi[1], Denis Murphy[4], Leslie Low Eng Ti[1*]

1 Advanced Biotechnology and Breeding Centre, Malaysian Palm Oil Board, P.O. Box 10620, 50720 Kuala Lumpur, Malaysia; 2 Faculty of Science and Technology, Universiti Kebangsaan Malaysia, 43600 Bangi, Selangor, Malaysia; 3 Spatial Sciences Institute, University of Southern California, Los Angeles, CA, 90089, USA; 4 Genomics and Computational Biology research group, University of South Wales, Pontypridd, CF371DL, United Kingdom; 5 Softberry Inc., 116 Radio Circle, Suite 400 Mount Kisco, NY 10549, USA.

* To whom correspondence should be addressed. Tel. +60387694504. Fax. +60389261995. E-mail: lowengti@mpob.gov.my

[#] These authors contributed equally to the paper and should be considered joint first authors.

**Abstract**

The advent of rapid and inexpensive DNA sequencing has led to an explosion of data that must be transformed into knowledge about genome organization and function. Gene prediction is customarily the starting point for genome analysis. This paper presents a bioinformatics study of the oil palm genome, including a comparative genomics analysis, database and tools development, and mining of biological data for genes of interest. We annotated 26,087 oil palm genes integrated from two gene-prediction pipelines, Fgenesh++ and Seqping. As case studies, we conducted comprehensive investigations on intronless, resistance and fatty acid biosynthesis genes, and demonstrated that the current gene prediction set is of high quality. 3,672 intronless genes were identified in the oil palm genome, an important resource for evolutionary study. Further scrutiny of the oil palm genes revealed 210 candidate resistance genes involved in pathogen defense. Fatty acids have diverse applications ranging from food to industrial feedstock, and we identified 42 key genes involved in fatty-acid biosynthesis in oil palm mesocarp and kernel. These results provide an important resource for studies on plant genomes and a theoretical foundation for marker-assisted breeding of oil palm and related crops.

# 1. Introduction

Oil palm belongs to the genus *Elaeis* of the family Arecaceae. The genus has two species - *E. guineensis* (African oil palm) and *E. oleifera* (American oil palm). *E. guineensis* has three fruit forms that mainly vary in the thickness of their seed (or kernel) shell - *dura* (thick-shell), *tenera* (thin-shell) and *pisifera* (no shell). The African oil palm is by far the most productive oil crop[1] in the world, with estimated production in year 2015/2016 of 61.68 million tonnes, of which the Malaysian share was 19.50 million tonnes[2]. Palm oil constitutes ~34.35% of the world production of edible oils and fats. Globally, palm oil is mainly produced from *E. guineensis,* in the *tenera* form. *E. oleifera,* is little planted because of its low yield (only 10 - 20% of *guineensis*). However, it is more disease-resistant and planted in areas where *guineensis* is well-nigh impossible, e.g., Central-Southern America. Even then, it is mainly planted as a backcross to *guineensis* (interspecific hybrid) to raise its yield. Nevertheless, it has economically valuable traits which plant breeders drool over to introgress into *guineensis*, such as a more liquid oil with higher carotenoid and vitamin E contents, disease resistance and slow height increment.[1]

The importance of oil palm has resulted in considerable interest to sequence its transcriptomes and genome. Initial work used expressed sequence tags (ESTs)[3], a technique very useful for tagging expressed genes but only providing partial coverage of the coding regions and genome in general. Next, GeneThresher$^{TM}$ technology was applied to selectively sequence hypomethylated regions of the genome.[4] With the genome sequenced[5], then using genetic mapping and homozygosity mapping via sequencing, the *SHELL* gene was identified.[6] This discovery allowed for an efficient genetic test to distinguish between the *dura, pisifera* and *tenera* fruit forms. Subsequently, the *VIRESCENS* gene, which regulates the fruit exocarp colour[7], and the *MANTLED* gene, which causes tissue culture abnormality[8], were also discovered. Accurate genome annotation was critical for the identification of these genes, and will be crucial for increasing the palm productivity.

*de novo* gene prediction was first established in the 1990s from modelling. A few years later, the Genscan[9] software predicted multiple and partial genes on both strands, with improved accuracy. Then, a host of programmes just about exploded to navigate the complexity of various genomes. Combining multiple predictors led to the development of automated pipelines to integrate the various experimental evidences.[10] Currently, such pipelines are primarily used for annotating whole genomes. A major limitation of many of the current approaches is their relatively poor performance in organisms with atypical distribution of nucleotides.[11–14] Accurate gene prediction and the discovery of regulatory elements in promoter sequences are two of the most important challenges in computational biology, for the prediction quality affects all aspects of genomics analysis.

To overcome the lack of precision in many predictive models, we developed an integrated gene-finding framework, and applied it to identify high quality oil palm gene models. The framework is a combination of the Seqping[15] pipeline developed at MPOB, and the Fgenesh++[16] pipeline at Softberry. Parameters of the framework were optimized to reliably identify the genes. Individual components of the framework were trained on known genes of plants closely related to the oil palm, such as the date palm, to identify the most suitable parameters for gene prediction. We identified a high confidence set of 26,087 oil palm genes. This paper presents a bioinformatics analysis of the oil palm genome, including comparative genomics analysis, database and tools development, and characterization of the genes of interest.

## 2. Materials and Methods

### 2.1 Datasets

We used the *E. guineensis* P5-build scaffold assembly from Singh et al.,[5] which contained 40,360 genomic scaffolds. The *E. guineensis* mRNA dataset is a compilation of published transcriptomic sequences from Bourgis et al.,[17] Tranbarger et al.,[18] Shearman et al.,[19,20] and Singh et al.,[6] as well as 24 tissue-specific RNA sequencing assemblies from MPOB, which were submitted to GenBank in BioProject PRJNA201497 and PRJNA345530, and expressed sequence tags (ESTs) downloaded from the nucleotide database in GenBank. The dataset was used to train the Hidden Markov Model (HMM) for gene prediction, and as transcriptome evidence support.

### 2.2 Fgenesh++ Gene Prediction

Fgenesh++ (Find genes using Hidden Markov Models)[16,21] is an automatic gene prediction pipeline, based on Fgenesh, a HMM-based *ab initio* gene prediction program.[22] We used the P5-build genomic scaffolds to make the initial gene prediction set using Fgenesh gene finder and generic parameters for monocot plants. From this set, we selected a subset of predicted genes that encode highly homologous proteins (using BLAST) to known plant proteins from the NCBI non-redundant (NR) database. Based on this subset, we computed gene-finding parameters specific to oil palm, and executed the Fgenesh++ pipeline to annotate the genes in our genomic scaffolds. The Fgenesh++ pipeline takes into account all the available supporting data, such as known transcripts and homologous protein sequences. To predict genes, we compiled a subset of NR plant transcripts mapped to genomic sequences providing a set of potential splice sites to the Fgenesh++pipeline. Also, the pipeline (by its *Prot_map* module) aligned all proteins from a compiled subset of NR database with plant proteins to the genomic contigs and found highly homologous proteins in gene identification. All predicted amino acid sequences were compared to protein sequences in the NR database, using BLAST (E-value cutoff: 1E-10). BLAST analysis of the predicted sequences was also carried out against the *E. guineensis* mRNA dataset, using an identify cutoff of >90%.

### 2.3 Seqping Gene Prediction

Seqping[15], a customized pipeline based on MAKER2,[23] was developed by MPOB. Full-length open reading frames (ORFs) were identified from the *E. guineensis* mRNA dataset described above, using the EMBOSS *getorf* program. The range of ORF lengths selected was 500 - 5000 nt, to minimize potential prediction errors. Using BLASTX[24] search, ORFs with E-values <1E-10 were deemed significantly similar to the RefSeq plant protein sequences. ORFs with BLASTX support were clustered using BLASTClust and CD-HIT-EST[25], and subsequently filtered using the TIGR plant repeat database,[26] GIRI Repbase[27] and Gypsy Database[28] to remove ORFs similar to retroelements. The resultant ORFs were used as the training set to develop the HMMs for GlimmerHMM,[29,30] AUGUSTUS[31] and SNAP,[32] which were subsequently used for gene predictions. Seqping used MAKER2[23] to combine the predictions from the three models. The predicted sequences were compared to the RefSeq[33] protein sequences and *E. guineensis* mRNA dataset via BLAST analysis.

**2.4    Integration of Fgenesh++ and Seqping Gene Predictions**

To increase the annotation accuracy, genes predicted by the Seqping and Fgenesh++ pipelines were combined in a unified gene prediction pipeline. ORF predictions with <300 nucleotides were excluded. Predicted genes from both pipelines in the same strand were considered overlapping if the overlap length was above the threshold fraction of the shorter gene length. A co-located group of genes was considered to constitute a locus if every gene in the group overlaps at least one other member of the same group (a single linkage approach). Different overlap thresholds, from 60% to 95% in 5% increments, were tested to determine the best threshold for further analysis, simultaneously maximizing the annotation accuracy and minimizing the number of single-isoform loci.

Gene functions were determined using PFAM-A[34,35] (release 27.0) and PfamScan ver. 1.5. The palm coding sequences (CDSs) were also compared to the NR plant sequences from RefSeq (release 67), using *phmmer* function from the HMMER-3.0 package.[36,37] To find the representative isoform and function for each locus, we took the lowest E-value isoform in each locus and the function of its RefSeq match. We excluded hits with E-values >1E-10, keeping only high-quality loci and the corresponding isoforms. The ORF in each locus with the best match to the RefSeq database of all plant species was selected as the *best representative* CDS for the locus. Gene ontology (GO) annotations were assigned to the palm genes, using the best NCBI BLASTP hit to match *O. sativa* sequences from the MSU rice database[38] at a cutoff E-value of 1E-10.

**2.5    Intronless Genes**

Intronless genes (IG) were identified through a set of mono-exonic genes containing full-length ORFs, as specified by the gene prediction pipeline. The same approach was applied to five other genomes: *A. thaliana* (TAIR10),[39] *O. sativa* (MSU 6.0),[38] *S. bicolor* (Phytozome 6.0), *Z. mays* (Phytozome) and *Volvox carteri* (Phytozome 8.0).[40] Lists of non-redundant IG from all six genomes were obtained, and the oil palm IG compared with those in the other genomes using BLASTP (E-value cutoff: 1E-5). The protein sequences of the IG were also mapped to all NCBI genes in the Archaea, Bacteria and Eukaryote kingdoms using BLASTP with the same cutoff.

**2.6    Resistance (R) Genes**

All curated plant resistance (R) genes were downloaded from the database PRGdb 2.0.[41] A local similarity search of known plant resistance genes and oil palm gene models was done using the BLASTP program with E-value ≤ 1E-5. TMHMM2.0[42] was used to find predicted transmembrane helices in the known R genes, as well as in the oil palm candidate R genes, and these results were used for classification of the R genes. Domain structures of the known and oil palm candidate R genes were identified using InterProScan. All domains found in the known R genes were used to classify candidate R genes in oil palm according to the PRGdb classification. To be considered an R gene, a candidate was required to contain all the domains found in known R genes of a certain class. Our selection was validated based on the published "resistance" gene motifs[43–47] and each class validated via multiple sequence alignment and phylogenetic tree, using the ClustalW[48] and MEGA6[49] programs, respectively. The same procedure was used to identify R genes in the five other genomes for comparison.

## 3. Results and Discussion

### 3.1 Gene models

Prediction and annotation of protein-coding genes are routinely done using a number of software tools and pipelines, such as Fgenesh++,[16] MAKER-P,[77] Gramene,[78] and Ensembl.[79] In plants, at least three model organisms (*A. thaliana*, *Medicago truncatula* and *O. sativa*) have been annotated using a combination of evidence-based gene models and *ab initio* predictions.[80–82] The first oil palm genome[5] was published in 2013 with assembled sequences representing ~83% of the 1.8Gb-long genome. Gene models were predicted by integrating two pipelines: Fgenesh++ and Seqping[15] (MPOB's in-house tool).

Previous studies of five *ab initio* pipelines (Fgenesh++, GeneMark.hmm, GENSCAN, GlimmerR and Grail) to evaluate the precision in predicting maize genes showed that Fgenesh++ produced the most accurate annotations.[21] Fgenesh++ is a common tool for eukaryotic genome annotation, due to its superior ability to predict gene structure.[83–86] In the oil palm genome, Fgenesh++ predicted 117,832 whole and partial-length gene models at least 500 nt long. A total 23,028 Fgenesh++ gene models had significant similarities to the *E. guineensis* mRNA dataset and RefSeq proteins (Table 1).

To improve the coverage and accuracy of the predicted gene models, and to minimize prediction bias, Seqping, which is based on the MAKER2 pipeline[23], was used. The pipeline was customized to achieve high accuracy predictions for the oil palm genome. It identified 7,747 putative full-length CDS from the transcriptome data used to train the models GlimmerHMM,[29,30] AUGUSTUS,[31] and SNAP.[32] Oil palm-specific HMMs were used in MAKER2 to predict oil palm genes. The initial prediction identified 45,913 gene models that were repeat-filtered and scrutinized using the oil palm transcriptome dataset and RefSeq protein support (Table 1).

The gene models were combined to identify the number of independent genomic loci. At an 85% overlap threshold, we obtained 31,413 combined loci (Table 2 and Fig. 1). Next, we chose a subset of these loci that contained PFAM domains and RefSeq annotations, and got 26,087 high quality genomic loci. Of them, 9,943 contained one ORF, 12,163 two, and 3,981 three or more. For each locus, the ORF with the best match to plant proteins from the RefSeq database was selected as its *best representative* CDS. The analysis of gene structures showed that 14% were intronless and 16% contained only two exons. We identified 395 complex genes with 21 to 57 exons each. The distributions of exons per gene and mRNA and CDS lengths are shown in Figure 2A and Figure 2B, respectively. The highest number of genes was at the 501-1000 nt interval, with 34.40% of mRNA and 30.62% of ORFs. The longest ORF and mRNA were 14,505 and 14,685 nt, respectively. Benchmarking Universal Single-Copy Orthologs (BUSCO)[87] analysis of the representative gene models showed that 90.44% of the 429 eukaryotic BUSCO profiles and 94.04% of the 956 *plantae* BUSCO profiles were available, thus quantifying the completeness of the oil palm genome annotation.

### 3.2  Fraction of cytosines and guanines in third position of codon, GC$_3$

GC$_3$ is defined as $\frac{C_3+G_3}{(L/3)}$, where $L$ is the length of the coding region, $C_3$ the number of cytosines, and $G_3$ the number of guanines in the third position of codons in the coding region.[88] The frequency of guanine and cytosine contents in the third codon position is an important characteristic of a genome.[88] Two types of GC$_3$ distribution have been described - unimodal and bimodal.[88–90] Genes with high and low GC$_3$ peaks have distinct functional properties:[90] GC$_3$-rich genes provide more targets for methylation, exhibit more variable expression, more frequently possess upstream TATA boxes and are predominant in stress responsive genes. Different gene prediction programs are variously biased to different classes of genes, with GC$_3$ rich genes reportedly especially hard to predict.[91]

Figure 3A shows the distribution of GC$_3$ in the high quality dataset. We ranked all genes by their GC$_3$ content and designated the top 10% (2,608 ORFs) as GC$_3$-rich (GC$_3 \geq 0.7535$), and the bottom 10% as GC$_3$-poor (GC$_3 \leq 0.37$). Two of the remarkable features that distinguish GC$_3$-rich and -poor genes are the gradients of GC$_3$ and CG$_3$-skew (defined as $CG_3^{skew} = \frac{C_3-G_3}{C_3+G_3}$). The increase in $CG_3^{skew}$ from 5' to 3' has been linked to transcriptional efficiency and methylation status.[88,90,92] Figures 3C and D show a positional change in nucleotide composition. The GC$_3$ content of GC$_3$-rich genes increases from the 5' to 3' end of the gene, but decreases in GC$_3$-poor genes. In addition, the plot of CG$_3$ skew increases in the proportion of cytosine in the GC$_3$-rich group, consistent with the hypothesis of transcriptional optimization of GC$_3$-rich genes. Despite the small GC$_3$-rich and GC$_3$-poor datasets, we observed pronounced peaks of nucleotide frequencies near the predicted start of translation, as also found in other well-annotated genomes.[88]

The genome signature, defined as $\rho_{CG} = \frac{f_{CG}}{f_C f_G}$, where $f_x$ is the frequency of (di)nucleotide, $x$, shows the distribution of methylation targets. Similar to grasses, and other previously analyzed plant and animal species,[88,90] the palm genome signature differs for GC$_3$-rich and GC$_3$-poor genes (Fig. 3B). The GC$_3$-rich genes are enriched and the GC$_3$-poor genes are depleted in the number of CpG sites. Many of the GC$_3$-rich genes are stress-related, while many of the GC$_3$-poor genes have housekeeping functions (see GO annotation in Supplementary Table S1). The depletion of CpGs in GC$_3$-poor genes is consistent with their broad, constitutive expression.[88] Next, we calculated the distribution of all nucleotides in the coding regions. We considered the following models of ORF: *Multinomial* (all nucleotides independent, and their positions in the codon not important), *Multinomial* position-specific and *First order three periodic Markov Chain* (nucleotides depend on those preceding them in the sequence, and their position in the codon is considered). Supplementary Table S2-S5 shows the probabilities of nucleotides A, C, G and T in $GC_3$-rich and -poor gene classes. Notice that both methods predict that $GC_3$-poor genes have greater imbalance between C and G, than $GC_3$-rich genes (-18.1% vs. 29.4%). This is consistent with the prior observation[90] that $GC_3$-poor genes are more methylated than $GC_3$-poor genes, and some cytosine can be lost from this deamination.

$GC_3$-rich and -poor genes differ in their predicted lengths and open reading frames (Supplementary Table S6): the $GC_3$ rich genes have gene sequences and ORFs approximately seven times and two times shorter, respectively, than the $GC_3$-poor genes. This is consistent with the findings from other species.[88–90] It is important to note that $GC_3$-rich genes in plants tend to be intronless.[88]

## 3.3 Intronless Genes (IG)

Intronless genes (IG) are common in single-celled eukaryotes, but are only a small percentage of the genes in metazoans.[93,94] Across multi-cellular eukaryotes, IG are frequently tissue- or stress-specific, $GC_3$-rich and their promoters have a canonical TATA-box.[88,90,93] Among the 26,087 representative gene models, 3,672 (14.1%) were IG. The mean $GC_3$ content of IG is 0.668±0.005 (Fig. 4), while the intron-containing (a.k.a. multi-exonic) genes' mean $GC_3$ content is 0.511±0.002, in line with other species. Intronless CDS are, on average, shorter than multi-exonic CDS: 924 ±19 nt *vs*. 1289±12 nt long.

The distribution of IG in the whole genome is different for various functional groups.[88,94] For example, in the oil palm genome, 29% of the cell-signaling genes are intronless, compared to just 1% of all tropism-related genes (Supplementary Table S7). The distribution of genes by GO categories is similar to that in *O. sativa*. It has been shown that in humans, mutations in IG are associated with developmental disorders and cancer.[94] Intronless and $GC_3$-rich genes are considered evolutionarily recent[88] and lineage-specific,[93] potentially appearing as a result of retrotransposon activity.[94,95] It is reported that 8 - 17% of the genes in most animals are IG - ~10% in mice and humans[93] and 3 - 5% in teleost fish. Plants have proportionately more IG than animals: 20% in *O. sativa*, 22% in *A. thaliana*,[96] 22% in *S. bicolor*, 37% in *Z. mays*, 28% in foxtail millet, 26% in switchgrass and 24% in purple false brome.[97] We have independently calculated the fraction of IG in *O. sativa*, *A. thaliana*, *S. bicolor* and *Z. mays* using the currently published gene models for each species, with results of 26%, 20%, 23% and 37%, respectively. (Supplementary Table S8). To establish a reference point, we calculated the fraction of IG in the green algae, *V. carteri*, and found 15.8%. The high IG in grasses is not surprising, since they have a clearly bimodal distribution of $GC_3$ composition in their coding region, with the $GC_3$-peak of this distribution dominated by IG.[88]

Using BLASTP, we found 555 IG (15% of oil palm IG) conserved across all the three domains of life: Archaea, Bacteria, and Eukaryotes (Fig. 5). These genes are likely essential for their survival.[98] A total of 738 oil palm IG were only homologous to bacterial genes, while 40 IG were only shared with Archaea. This is consistent with the extreme growth conditions of extant Archaea from the typical conditions of Bacteria; meaning that there are fewer opportunities for horizontal gene transfer from Archaea than from Bacteria to the oil palm genome. Considering the three kingdoms in eukaryotes, we observed 1,384 oil palm IG shared among them. A significant portion of the oil palm IG (1,864) was only homologous to Viridiplantae (green plants). These proteins may have evolved, or been regained, only in plants, while other organisms have lost their ancestral genes during evolution.[96]

Reciprocal BLAST was carried out to verify homologies of the oil palm candidate IG to produce a set of high confidence oil palm IG. We found 2,433 (66.26%) proteins encoded by oil palm IG to have orthologs in *A. thaliana*, *O. sativa* or *Z. mays* that were also intronless, indicating that intronlessness is an ancestral state.[99,100] In conclusion, from our representative gene models, we estimate that about one-seventh of the genes in oil palm are intronless. We hope that this data will be a resource for further comparative and evolutionary analysis, and aid in the understanding of IG in plants and other eukaryotic genomes.

**Conclusions**

We developed an integrated gene prediction pipeline, enabling annotation of the genome of the African oil palm and present a set of 26,087 high quality and thoroughly validated gene models. To achieve this, we conducted an in-depth analysis of several important gene categories: IG, R and FA biosynthesis. The prevalence of these groups was similar across several plant genomes, including *A. thaliana*, *Z. mays*, *O. sativa, S. bicolor*, *G. max* and *R. communis*. Coding regions of the oil palm genome have a characteristic broad distribution of $GC_3$, with a heavy tail extending to high $GC_3$ values where it contains many stress-related and intronless genes. $GC_3$-rich genes in oil palm are significantly over-represented in the following GOslim process categories: responses to abiotic stimulus, responses to endogenous stimulus, RNA translation, and responses to stress. BUSCO analysis showed that our high-quality gene models contained at least 90% of the known conserved orthologs in eukaryotes. We also found that approximately one-seventh of the oil palm genes intronless. We detected a similar number of R genes in the oil palm compared to other sequenced plant genomes. Lipid-, especially fatty acid (FA)-related genes, are of interest in oil palm where, in addition to their roles in specifying oil yield and quality, they also contribute to the plant organization and function as well as having important signaling roles related to biotic and abiotic stresses. We identified 42 key genes involved in FA biosynthesis in oil palm. Our study will facilitate understanding of the plant genome organization, and will be an important resource for further comparative and evolutionary analysis. The study on these genes will facilitate further studies on the regulation of gene function in oil palm and provide a theoretical foundation for marker-assisted breeding of the crop with increased oil yield and elevated contents of oleic and other value-added fatty acids.

**Acknowledgements**

We thank Orion Genomics LLC for their assistance and advice in data analysis. Special thanks to the Breeding and Tissue Culture Unit for the supply of palms and RNA extraction for sequencing. Last, but not least, we extend our appreciation to the Director General MPOB, Dr. Ahmad Kushairi Din, for his support and encouragement throughout the project. Dr. Tatarinova was supported by NSF Awards #1456634 and #1622840.